\documentclass[iop]{emulateapj}

\usepackage{epsfig}
\usepackage{amsmath}
\usepackage{natbib}

\newcommand{\be}{\begin{equation}}
\newcommand{\ee}{\end{equation}}
\newcommand{\ba}{\begin{eqnarray}}
\newcommand{\ea}{\end{eqnarray}}
\newcommand{\bfi}{\begin{figure}
\epsfxsize=9cm
\epsffile}
\newcommand{\efi}{\end{figure}}
\newcommand{\no}{\nonumber}

\begin{document}
\title{Diagnosing multiplicative error by lensing
  magnification of type Ia supernovae}
\author{Pengjie Zhang}
\affil{Center for Astronomy and Astrophysics, Department of
  Physics and Astronomy, Shanghai Jiao Tong University, Shanghai,
  China}
\affil{IFSA Collaborative Innovation Center, Shanghai Jiao Tong
University, Shanghai 200240, China}
\affil{Key Laboratory for Research in Galaxies and Cosmology, Shanghai
  Astronomical Observatory, 80 Nandan Road, Shanghai, China}
\email{zhangpj@sjtu.edu.cn}
\begin{abstract}
Weak lensing causes spatially coherent fluctuations in flux of type Ia supernovae (SNe
Ia). This lensing magnification allows for weak lensing measurement
independent of cosmic shear. It is free of shape measurement errors
associated with cosmic shear and can therefore be used to diagnose and
calibrate multiplicative error. Although 
this lensing magnification is difficult to measure accurately in auto correlation, its cross correlation
with cosmic shear and galaxy distribution in overlapping area can be
measured to significantly higher
accuracy. Therefore these cross correlations can put useful constraint on multiplicative
error, {\it and the obtained constraint  is free of cosmic variance in weak lensing  field}. We
present two methods implementing this idea and estimate their 
performances. We find that, with $\sim 1$ million SNe Ia that can be
achieved by the proposed D2k survey with the LSST telescope \citep{Zhan08}, 
multiplicative error of $\sim 0.5\%$ for source galaxies at $z_s\sim 1$
can be detected and larger multiplicative error can be corrected to the
level of $0.5\%$.  It is therefore a promising approach to control the
multiplicative to the sub-percent level required for stage IV
projects. The combination of the two methods even has the
potential to diagnose and calibrate galaxy intrinsic alignment, which
is another major systematic error in cosmic shear cosmology. 
\end{abstract}
\maketitle

\section{Introduction}
Weak gravitational lensing has great potential of probing dark matter, neutrinos,
dark energy and gravity at cosmological scales
\citep{Refregier03,DETF,Hoekstra08,Munshi08,Weinberg13}. All these great
applications rely on  accurate weak
lensing measurement. Cosmic shear, lensing induced coherent distortion in galaxy
shapes,  can achieve sub-percent statistical error in weak lensing
measurement. It is therefore a major science driver for 
massive cosmological surveys such as DES, Euclid, HSC, LSST and
SKA radio survey. However, cosmic shear suffers from a variety of systematic
errors such as photometric redshift errors and galaxy intrinsic
alignment (for reviews, refer to
\citet{LSST,LSSTDE12,Troxel14}). Tremendous efforts have been put to 
understand and correct these systematic errors. 

One systematic error
which has received intensive scrutiny is shear estimation error. It is
often conveniently decomposed into a 
multiplicative error and an additive error \citep{STEP1}. Stage IV
projects such as Euclid and LSST put stringent requirement of controlling multiplicative error
to $\sim 0.2\%$-$0.5\%$ \citep{Huterer06,Cropper13,Massey13}. A series of
blind community challenges of massive scale have been carried out over the last
decade (STEP1: \citet{STEP1}; STEP2: \citet{STEP2}; GREAT08:
\citet{GREAT08}; GREAT10: \citet{GREAT10}; GREAT3:
\cite{GREAT3}). The
latest GREAT3 challenge shows that control over multiplicative
  error/bias has been significantly improved. For mock catalogues,
  various shear estimation methods can control multiplicative error to
  $1\%$ or even $0.1\%$ when PSF is given \citep{GREAT3}.  Further improvement may still be 
expected by refining existing shear estimation methods or emerging new
methods such as the recently proposed  Fourier-space method \citep{ZhangJ08,ZhangJ10,ZhangJ11a,ZhangJ11b,ZhangJ13}. 

Nevertheless, the performance of shear estimation
methods depends on many factors such as galaxy size, flux (S/N), morphology, selection
criteria, weighting scheme, and the accuracy of PSF interpolation (e.g. GREAT3:
\citet{GREAT3}).  Given unprecedented variety of galaxies  at $z\sim
0-4$ of stage IV surveys, one must keep caution on whether these shear 
estimation methods can achieve the accuracy estimated from simulated lensed galaxies. 
It would then be safer to design and apply
independent diagnostics of multiplicative bias based on real data in a
model-independent manner. If multiplicative bias is detected by such
diagnostics, it can then be calibrated consequently. \citet{Vallinotto11} proposed to calibrate the multiplicative 
error against lensing magnification in galaxy flux and size, and
demonstrated the potential of such diagnostics. In this paper, we
propose an alternative method, that is to calibrate multiplicative
error by lensing magnification of type Ia supernovae (SNe Ia). 
 
Supernova (SN) flux is magnified by gravitational lensing by a factor
$\mu\equiv 1/[(1-\kappa)^2-\gamma^2]$. Here, $\mu$, $\kappa$ and
$\gamma_{1,2}$ are the lensing magnification/amplification, convergence and shear,
respectively. $\gamma^2\equiv \gamma_1^2+\gamma_2^2$. In the weak
lensing regime, the measured flux fluctuation of SNe Ia (after standalization) is 
$\delta_F=\mu-1+\delta_F^{\rm int}$. Here $\delta_F^{\rm int}$ is the
intrinsic flux fluctuation of SNe Ia. On
one hand, this lensing magnification contaminates the Hubble diagram
and degrades cosmological constraints from SNe Ia distance-redshift
measurement \citep{Kantowski95,Frieman96,Holz98,Dalal03,Cooray06b}. On the other hand, it 
provides independent measure of weak lensing through the lensing induced flux fluctuation 
\citep{Metcalf99,Hamana00,Menard05,Cooray06a,Dodelson06,Zentner09,Ben-Dayan13,Amendola14,Fedeli14},
and is therefore a useful source of information. Existing data
  already allows for marginal detection of lensing magnification in SN flux
\citep{Kronborg10,Betoule14,Castro14}.  With orders of
magnitude more $z\sim 1$ SNe Ia expected in future surveys, precision lensing measurement
through SN magnification is very promising. 

Lensing measured in this way is free of multiplicative error
troubling cosmic shear measurement. The measured cosmic 
shear $\gamma_{i=1,2}$ can be conveniently parametrized as
\citep{STEP1}, 
\be
\gamma_i^{\rm obs}=(1+m_i)\gamma_i+c_i+\gamma_i^{\rm int}\ ,
\ee
with an extra term $\gamma_i^{\rm int}$ arising from
the intrinsic galaxy shape noise. For the moment we approximate the measured
$\gamma/(1-\kappa)$ (reduced shear) as
shear $\gamma$. The neglected complexity will be discussed in \S
  \ref{sec:discussion}. $\gamma_i^{\rm int}$ in general has a dominant
  component of no spatial correlation, and  a spatially correlated component
(galaxy intrinsic alignment). $m_i$ is the multiplicative error/bias
and $c_i$ is the additive error. The two $m_i$ can differ form each
other (e.g. \cite{GREAT3}). For brevity, we will work with $m$
defined with respect to cosmic shear E-mode ($\kappa$). 

The two data sets (cosmic shear and SNe Ia magnification) both
measure the same weak lensing, but with different prefactors
(e.g. $1+m$). Therefore combining the two data sets we can measure $m$
without assumptions on the true lensing signal. Furthermore, if the
two data sets locate in the same cosmic volume,  cosmic
variance of the weak lensing field will be eliminated and will not
degrade constraint on $m$. 

The major obstacle in this approach is the low
number density of SNe Ia and therefore heavy shot noise. Later in this
paper we will show that at least half a million SNe Ia are required to
diagnose $|m|\la 1\%$,  the minimum requirement for stage IV
weak lensing surveys \citep{Huterer06}. Surveys of a million SNe Ia with well measured
light curves are highly ambitious. Nevertheless, surveys of such scale
can be accessible by telescopes like the LSST telescope and have been
proposed \citep{Zhan08,LSSTDE12}.  Cosmological benefits of such surveys will be
many-folds, besides the luminosity distance measurement and  peculiar
velocity measurement.  The large scale structure of these SNe Ia allows for measurement of baryon acoustic oscillation, which can significantly  
improve cosmological constraints from weak lensing alone
\citep{Zhan08}.  The proposed diagnosis of $m$ is a new bonus of such
supernova survey.

To quantify the capability of diagnosing and calibrating
multiplicative bias with SN magnification, we adopt the baseline survey of SNe Ia as the D2k survey proposed in 
\citet{Zhan08}. This proposed five year survey over $2000$ deg$^2$ by the LSST
telescope will result in about 2 million SNe Ia with well measured
light curves. Among them,  $0.7$ million locate at
$0.8<z<1.2$, one of the primary target redshift bins for precision
weak lensing measurement. For LSST cosmic shear, we assume a total of 3 billion
galaxies over 20000 deg$^2$, with a normalized redshift distribution
$n_g(z)=z^2\exp(-z/z_*)/(2z_*^3)$ \citep{Huterer06,Zhan06} and
$z_*=0.4$. The median redshift
is $2.675z_*=1.07$. The forecasted constraint on multiplicative error
calibration is sensitive to SN survey parameters, but is very
insensitive to cosmic shear survey parameters. 

This paper is organized as follows. We discuss two implementations (M1
and M2) of
diagnosing and calibrating multiplicative error combining cosmic shear
and SN magnification in \S \ref{sec:M1} and \S \ref{sec:M2}, respectively. We
discuss and conclude in \S \ref{sec:discussion}. Some technical
details of calculation are presented in the appendix.

\section{Method one}
\label{sec:M1} 
Method one only uses the two measurements, namely cosmic shear and SN
magnification, to calibrate multiplicative error.  For theoretical estimation of the expected S/N, it is much more
convenient to work in Fourier space than in real space. The observable
will be $\delta_F(\ell)$ and $\gamma^{\rm obs}(\ell)$. $\ell$ is an
independent multipole mode. Since we are only able to measure lensing
magnification of SNe Ia at $\ell\la 1000$ (Fig. \ref{fig:cl}),  we can treat the lensing
field as Gaussian.  The corresponding Fisher matrix is \citep{Tegmark97}
\be
\label{eqn:Fab}
{\bf F}_{\alpha\beta}=\sum_\ell \frac{1}{2}{\rm Tr}\left[{\bf C}^{-1}({\bf
    \ell}){\bf C}_{,\alpha}({\bf \ell}) {\bf C}^{-1}({\bf
    \ell}){\bf C}_{,\beta}({\bf \ell})\right]\ .
\ee
Here, $_{,\alpha(\beta)}\equiv \partial/\partial
\lambda_{\alpha(\beta)}$ and $\lambda_{\alpha(\beta)}$ is the $\alpha(\beta)$-th parameter to
be constrained. ${\bf C}(\ell)$ is the covariance matrix for the given 
$\ell$ mode, 
\be                                                                                 
\label{eqn:C}                                                                      
{\bf C}(\ell)=\bordermatrix{ & & \cr                                                      
                &C^\mu (\ell)+N^\mu&(1+m)C^{\mu\gamma}(\ell)\cr                                                 
                &(1+m)C^{\mu\gamma}(\ell)&(1+m)^2
                C^{\gamma}(\ell)+N^{\gamma} \cr}\ .
\ee
$C^\mu$, $C^\gamma$ and $C^{\mu\gamma}$ are the angular power spectra
of $\mu$,  $\gamma$ (E-mode) and their cross power spectrum, respectively. $N^{\mu}=4\pi
f_{\rm sky}\sigma_F^2/N_{\rm SN}$ is the noise power spectrum in SNe
Ia magnification measurement. $f_{\rm sky}$ is the fractional sky
coverage of overlapping SN survey and cosmic shear survey. For the D2k
survey, $f_{\rm sky}=2000/(4\times 180^2/\pi)=4.8\%$. $\sigma_F\equiv
\sqrt{\langle \delta_F^{\rm int,2}\rangle}$ is the rms dispersion of
flux of  standardized SNe Ia.  $N_{\rm SN}$
is the total number of SNe Ia in the given sky area and in the given
redshift bin. $N^{\gamma}=4\pi f_{\rm
  sky}\sigma_\epsilon^2/N_\gamma$ is the noise power spectrum in
cosmic shear measurement, $\sigma_\epsilon$ is the r.m.s ellipticity, and $N_\gamma$ is the total number of
galaxies in the given sky area and redshift bin for cosmic shear measurement.  We
will take the approximation $\mu\simeq 1+2\kappa$. On the other hand,
the E-mode shear $\gamma_E=\kappa$. Therefore
$C^\mu=4C^\gamma$ and $C^{\mu\gamma}=2C^\gamma$. 

In numerical evaluation of $\sigma_m$ throughout the paper, we
adopt $\sigma_F=0.1$ and $\sigma_\epsilon=0.3$.  $\sigma_F$ quoted
here is solely the intrinsic scatter $\sigma_F^{\rm int}$. In reality, it should include that
  induced by photo-z error. Photo-z
  error of $\sigma_z$ increases SN scatter to $\sigma_F\simeq \sigma_F^{\rm
    int}[1+a(\sigma_z/\sigma_\mu^{\rm int})^2]$, with $a=2(d\ln
  D_L(z)/dz)^2=2.5$ at $z=1$. For $\sigma_F^{\rm int}=0.1$ and
  $\sigma_z=0.01(1+z)$ \citep{Zhan08}, we have $\sigma_F\simeq 1.1\sigma_F^{\rm
    int}=0.11$.  We then conclude that including photo-z error does
  not significantly change our forecast. We focus on
redshift bin $0.8<z<1.2$, in which $0.72\times 10^6$ SNe Ia can be
observed by the proposed D2k survey. Fig. \ref{fig:cl}
plots $C^\gamma$, $N^\mu/4$ and $N^\gamma$ for $0.8<z<1.2$.  The
lensing power spectrum is calculated using the Limber integral, in
which the nonlinear matter power spectrum is calculated
using the halofit model \citep{Smith03}.  Due to 
sparse SN samples, each single lensing multipole mode
is overwhelmmed by shot noise at $\ell>80$. However, with $2\ell \Delta \ell f_{\rm sky}$
modes for each bin of width $\Delta \ell$, we can beat down shot noise by a factor $\sqrt{2\ell \Delta
  \ell f_{\rm sky}}$. Therefore we can still measure the lensing power
spectrum through SN magnification with $S/N>3$ at $\ell \sim 1000$ for
$\Delta \ell/\ell=0.1$.

\bfi{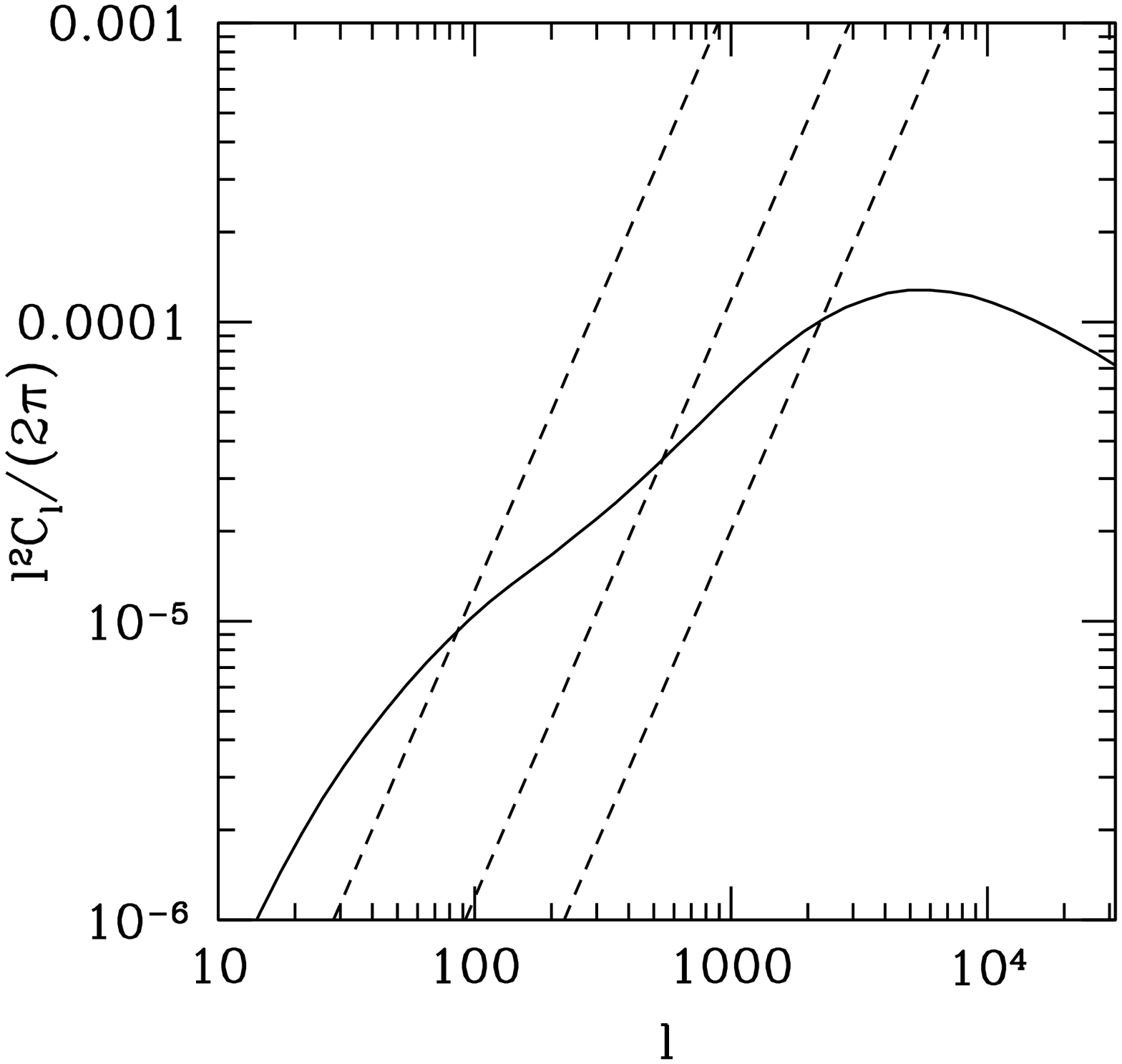}
\caption{The lensing power spectrum (solid line) and corresponding
  noises (dash lines). The source redshift is $0.8<z_s<1.2$. Shot noises, from large to small,
  are $N^\mu/4$ in SN magnification, $N^\gamma$ in cosmic shear and
  $N^{\rm gw}$ in the weighted galaxy distribution,
  respectively. Forecast of shot noise targets at the proposed D2k
  survey of $0.72$ million SNe Ia \citep{Zhan08} with the LSST
  telescope.  \label{fig:cl}}
\efi

\bfi{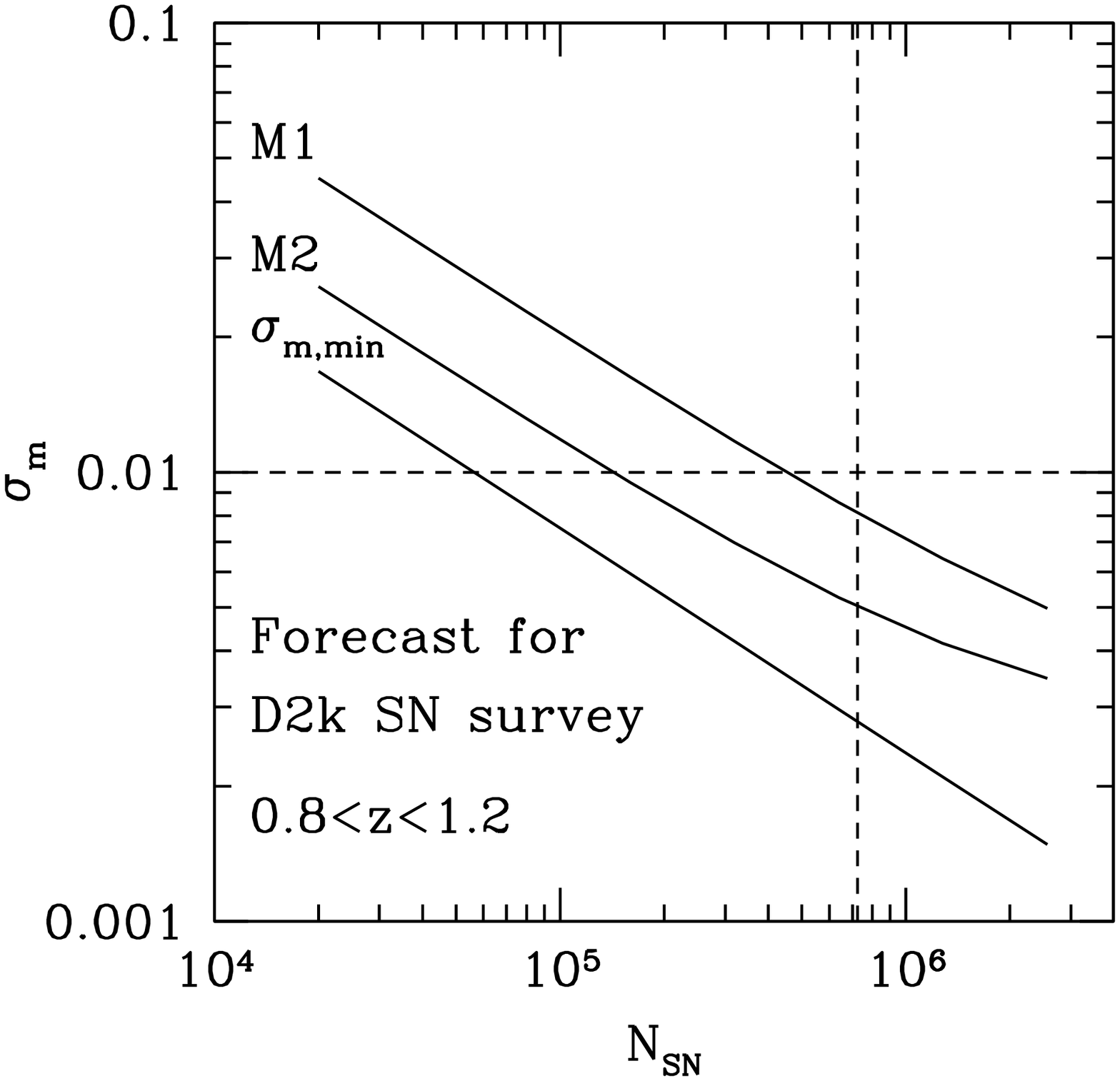}
\caption{The $1\sigma$ constraint on $m$ at $0.8<z<1.2$ as a function of the number
  of SNe Ia for the two methods (M1 and M2), and $\sigma_{m,{\rm
      min}}$, the lower limit of calibrating multiplicative error with SN
  magnification. The vertical dashed line denote $N_{\rm
    SN}=0.72\times 10^6$ of the proposed D2k survey with the LSST
  telescope \citep{Zhan08}. \label{fig:m}}
\efi

In the Fisher matrix analysis, we combine all $n$ independent
$\ell$ modes, which we label as $\ell_i$ ($i=1,\cdots,n$). We take  the unknown parameters to be $\lambda\equiv
(\lambda_0,\lambda_1,\cdots,\lambda_n)=(m,C^\gamma(\ell_1),\cdots,C^\gamma(\ell_n))$
with $\lambda_0=m$. The Fisher matrix ${\bf F}_{\alpha\beta}$ is calculated and inverted
in the appendix. We find that the error in $m$ is 
\ba
\label{eqn:m1}
\sigma_m\simeq \left(\int 
  \frac{2\ell d\ell f_{\rm sky} C^{\gamma,2}(\ell_i)}{C^\gamma(N^\mu/4+N^\gamma)+N^\gamma
    N^\mu/4}\right)^{-1/2} \ .
\ea
Immediately we find a fundamental lower limit for the $m$ calibration,
\ba
\label{eqn:mlimit}
\sigma_m>\sigma_{m, {\rm min}}&=&\frac{\sigma_F}{2\sigma_\kappa}N^{-1/2}_{\rm
  SN}\\
&=&5\times 10^{-3}
\left(\frac{\sigma_F}{0.1}\right)\left(\frac{0.01}{\sigma_\kappa}\right)\left(\frac{N_{\rm
    SN}}{10^6}\right)^{-1/2}\ . \no
\ea
Notice that $\sigma_\kappa^2\equiv \langle \kappa^2\rangle=\int
(\ell^2C^\gamma(\ell)/(2\pi))d\ell/\ell$. This fundamental lower limit corresponds to the limit that all other sources of statistical errors vanish and the only
one left is shot noise in SN magnification. 

This limit can only be achieved under the condition $N^\gamma\ll N^\mu$ and $N^\gamma\ll
C^\gamma$. The first condition is usually satisfied
(Fig. \ref{fig:cl}) since the galaxy population is much denser
that the SN population. For example,  the number density of cosmic shear
galaxies in a LSST-like survey at $0.8<z<1.2$ is $~600$ times higher
than that of SNe Ia even for an ambitious D2k survey, resulting in
$N^\gamma\sim 0.1 N^\mu/4$. It is for this reason that the constraint on the
multiplicative error $m$ is limited by SN survey configurations.

On the other hand, the second  condition $N^\gamma\ll C^\gamma$ is violated at $\ell\ga 500$ (Fig. \ref{fig:cl}), reflecting non-negligible
shot noise per multipole mode in cosmic shear measurement. Therefore in reality we are
not to reach the limit $\sigma_{m,{\rm min}}$

Fig. \ref{fig:m} shows $\sigma_m$ as a function of $N_{\rm SN}$. We
find that the actual $\sigma_m\sim 3\sigma_{m,{\rm min}}$
(Fig. \ref{fig:m}).  Nevertheless, the calibration accuracy on $m$ can reach $\sigma_m=8\times
10^{-3}$ for $N_{\rm SN}=0.72\times 10^6$ SNe Ia expected in the D2k
survey. This constraint is close to the requirement of $0.5\%$
  on $m$ for LSST \citep{Huterer06}  and is
  therefore encouraging.  Constraints of $m$ for other redshift bins
are shown in Table \ref{table:m}. 

Are there possibilities to further improve constraint on $m$?
Fig. \ref{fig:w} shows $({\rm S}/{\rm N})^2_\ell$, the constraining power per logarithmic $\ell$
bin defined through
\be
 \left(\frac{{\rm S}}{\rm N}\right)^2=\int \left(\frac{{\rm S}}{\rm N}\right)^2_ \ell\frac{d\ell}{\ell}\ . 
\ee
For the method discussed in this section (M1), the constraining
power peaks at $\ell\sim 1000$ (Fig. \ref{fig:w}).  Contribution from $\ell>1000$ is 
suppressed, since $N^\gamma\gg C^\gamma$ at $\ell>1000$
(Fig. \ref{fig:cl}).  Since most contribution 
  comes from relatively large scale $\ell \la 10^3$,  the Gaussian approximation adopted
  through the Fisher matrix estimation is valid.  However, Fig. \ref{fig:cl} shows that the lensing signal
peaks at $\ell\sim 3000$, so there are rooms for further
improvement. We present the second method (M2) to do so. 

\section{Method two}
\label{sec:M2}
Method two combines the galaxy distribution available in the same
survey, together with cosmic shear and SN magnification, to improve
constraint on multiplicative error. 
For a survey like LSST, the highest S/N measurement is for the galaxy
clustering (Fig. \ref{fig:cl}). Next is cosmic shear. SN magnification has the lowest
S/N. Therefore we can utilize the galaxy-SN magnification cross
correlation and galaxy-cosmic shear cross correlation to improve the
magnification and cosmic shear measurement. Combining the two cross
correlations allows better determination of $m$.

\bfi{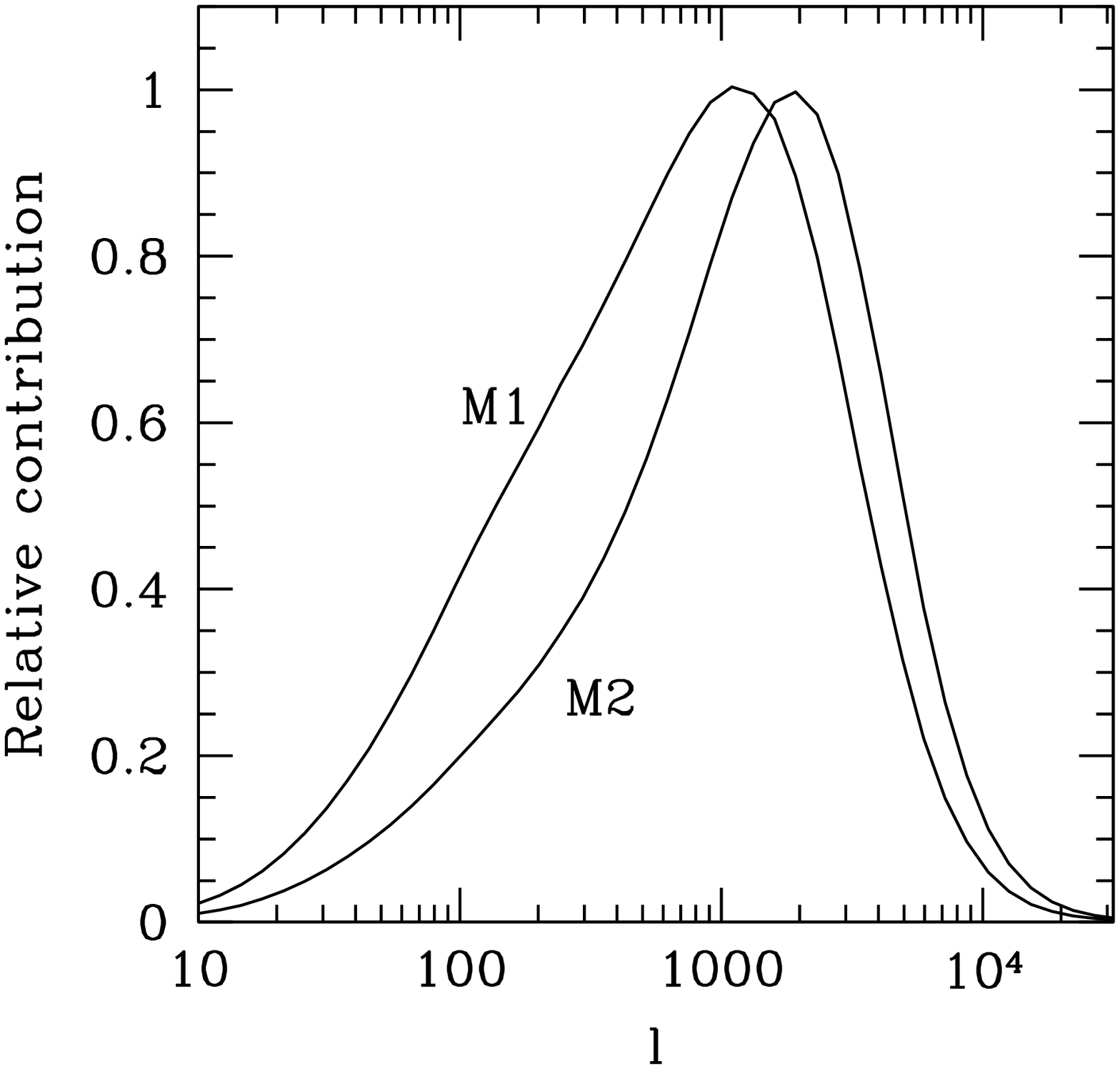}
\caption{Relative contribution per logarithmic $\ell$ bin for
  $0.8<z<1.2$ of the proposed D2k SN survey, for the two methods respectively. \label{fig:w}}
\efi

The galaxy surface overdensity is 
\be
\delta \Sigma_g=\frac{\int
  \delta_g n_g(z)W_g(z)dz}{\int n_g(z)W_g(z)dz}\ .
\ee
 $n_g(z)$ is the mean galaxy  redshift distribution and the redshift integral
 is over the given redshift bin. Since galaxies in LSST have photo-z
  information, we can apply a redshift dependent weighting $W_g(z)$  to improve the measurement accuracy of $m$.  We
  then have two measures of cross power spectra. $\hat{C}^{\mu g}$ is
  the measured  SN magnification-galaxy overdensity cross power
  spectrum. $\hat{C}^{\gamma g}$ is the measured cosmic shear-galaxy
  overdensity cross power spectrum. The ``hat'' on top of
  corresponding property (e.g. $\hat{C}$) denotes the measured
  quantity with measurement errors. We expect that $C^{\gamma
    g}=(1+m)C^{\mu g}/2$. Therefore, we can estimate $m$  combining the two measurements
  $\hat{C}^{\mu g}$ and $\hat{C}^{\gamma g}$, 
\be
\hat{m}=1-\frac{\hat{C}^{\mu g}(\ell)/2}{\hat{C}^{\gamma
    g}(\ell)}\ ,
\ee
The requirement here is that the bin size $\Delta \ell$ is sufficiently large so
error in $\hat{C}^{\gamma g}(\ell)$ is small ($\delta C^{\gamma g}\ll
C^{\gamma g}$).  The expectation value $\langle \hat{m}\rangle=m+O(m^2)$. Since $|m|\ll
1$, the above estimator is virtually free of systematic
bias. When taking the ratio, cosmic variance in the galaxy-shear
correlation cancels that in the galaxy-magnification correlation
because the two share identical cosmic volume and hence identical
cosmic variance . Constraint on $m$ from a single $\ell$ bin, assuming Gaussianity, is 
\ba
\label{eqn:mell}
\sigma^2_m(\ell)&=&\frac{(C^g(\ell)+N^{\rm gw})(N^\mu/4+N^\gamma)}{2\ell\Delta
  \ell f_{\rm sky} C^{\mu g,2}(\ell)/4}\\ 
&=&\frac{1}{2\ell\Delta
  \ell f_{\rm sky} r^2} \left(\frac{C^g+N^{\rm
    gw}}{C^g}\right)\left(\frac{N^\mu/4+N^\gamma}{C^\mu/4}\right)\no \ .
\ea
Here, $N^{\rm gw}=4\pi f_{\rm sky}/N^W_g$ is the shot noise in the weighted galaxy
clustering. $N_g^W=N_{\rm tot}/\langle W_g\rangle$ is the weighted
galaxy number in the given cosmic volume, and
$\langle W_g\rangle\equiv \int n_g(z)W_g(z)dz/\int n_g(z)dz$. Finally
we will combine all multipole bins to constraint $m$, 
\be
\sigma_m=\left[\sum_\ell
  \sigma_m^{-2}(\ell)\right]^{-1/2}\ .
\ee

In Eq. \ref{eqn:mell}, $r\equiv
C^{\mu g}/\sqrt{C^\mu C^g}$ is the cross correlation  
coefficient between the weighted galaxy distribution and lensing. An
important step to reduce the calibration error is to 
increase $r$. Due to the large amount of galaxies and strong
clustering between them, $C^g\gg N^{\rm gw}$ at $\ell \la 10^3$. We
then have the luxury to weigh
these galaxies to increase $r$.  Since we have (photometric)
redshift information of galaxies and we have accurate measurement of
galaxy bias, we can exert a nearly optimal weighting 
to galaxies such that their mean redshift distribution matches
that of the lensing kernel. The weighting is 
\be
W_g(z)=\frac{W_L(z)H_0}{n_g(z)b_g(z)H(z)}\ .
\ee
Here, $W_L$ is the lensing kernel defined through
\be
\kappa=\int \delta_m
  W_L(z)\frac{d\chi}{c/H_0}\ .
\ee
 $W_L(z)=\langle W_L(z,z_s)\rangle$ is 
  the  lensing kernel averaged over the source galaxy distribution.
  $W_L(z,z_s)$ is the lensing kernel of a single source redshift
  $z_s$ (e.g. \citet{Refregier03})
\begin{equation*}
W_L(z,z_s)=\begin{cases}
\frac{3}{2}\Omega_m(1+z)\frac{\chi(z)}{c/H_0}\left[1-\frac{\chi(z)}{\chi(z_s)}\right]& {\rm if}\ z<z_s\
,\\
0                  & {\rm if}\ z\geq z_s\ .
\end{cases}
\end{equation*}
 After this weighting, we expect $r\simeq 1$. Under this
  limit, the weighted galaxy angular power spectrum
  $C^g=C^\gamma=C^\kappa$.   The requirement to
    achieve $r=1$ is negligible stochasticity in galaxy
    bias. galaxy stochasticity will bring $r<1$ and
    therefore degrade constraint on $m$.  Fig. \ref{fig:w} shows that
    most constraining power comes from   $\ell\la 2000$, where stochasticity in galaxy distribution
    is insignificant. Therefore we only expect modest degradation in
    constraint of $m$ caused by stochasticity. Given complexities in
    modelling galaxy stochasticity, we will simply ignore it in this
    paper and only caution that $\sigma_m$ of method two can be slightly underestimated.

Under the condition $C^g\gg N^{\rm gw}$, we can prove that 
$\sigma_m=\sigma_{m,{\rm min}}$. However, Fig. \ref{fig:cl} shows that
even  this condition breaks at $\ell \ga 1000$, and even worse, $C^g\la N^{\rm
  gw}$ at $\ell\ga 2000$. Therefore the actual constraint on $m$ is
poorer ($\sigma_m>\sigma_{m,{\rm min}}$). Numerical results shown in
Fig. \ref{fig:m} find $\sigma_m\sim 1.5 \sigma_{m,{\rm
    min}}$ over a wide range $10^4<N_{\rm SN}<10^6$.  Therefore method two can deliver a factor of $2$ better constraint
on $m$, with respect to method one ($3\sigma_{m,{\rm min}}\rightarrow
1.5 \sigma_{m,{\rm min}}$). For $0.8<z<1.2$,
the constraining power peaks at $\ell
\sim 2000$, so method two utilizes more lensing information than
method one whose constraining power peaks at $\ell \sim 1000$
(Fig. \ref{fig:w}).  This explains why method two works better than
method one. 

With $0.72\times 10^6$ SNe Ia at $0.8<z<1.2$
that the proposed D2k survey can measure, method two can achieve
$\sigma_m=5\times 10^{-3}$. This basically meets the requirement for
LSST \citep{Huterer06}. Apply method two to other
redshift bins also turns out excellent constraints on $m$
($0.4$-$0.9\%$, Table \ref{table:m}). For this purpose, adding a SN
survey like the proposed D2k to LSST would be highly
beneficial.\footnote{Euclid requires $|m|\leq
0.2\%$ \citep{Cropper13,Massey13}. It would require a dedicated SN
survey (if any), with  twice as many SNe Ia as the D2k survey, to
meet the Euclid requirement.  }

\begin{table}
\caption{Sensitivity to multiplicative error for various redshift
  bins. The labels ``M1'' and ``M2'' denote the two calibration
  methods. The number of SNe Ia is based on the D2k survey proposed in
\cite{Zhan08}, which plans to cover 2000 deg$^2$ over five years with
the LSST telescope. The estimation adopts $\sigma_F=0.1$ and results for
other values of $\sigma_F$ should scale by $\sigma_F/0.1$. }
\begin{tabular}{rccccccccc} \hline\\
$[z_{\rm min},z_{\rm max}]$ & $[0.6,0.8]$ & $[0.8,1.0]$ & $[1.0,1.2]$ &
$[0.8,1.2]$ & $[0.6,1.2]$\\
$N_{\rm SN}$ & 0.59M  &  0.50M &0.22M &0.72M & 1.31M \\\hline\\
$\sigma_m(M1)$& 1.8\% & 1.4\% & 1.6\% & 0.8\%& 0.6\%\\
$\sigma_m(M2)$& 0.9\% & 0.7\% &0.8\% &0.5\%& 0.4\%\\\\ \hline
\end{tabular}
\label{table:m}
\end{table}
 
\section{Discussion}
\label{sec:discussion}
Method two is superior to method one in many aspects. The constraints on $m$ of various redshift bins are shown in Table
\ref{table:m}. $\sigma_m$ of method two is usually a factor of
$1.5$-$2$ smaller than that of method one, showing that method two is
superior in statistical error. Furthermore, method two is unbiased to
the presence of galaxy intrinsic alignment and additive error in cosmic shear
measurement. Galaxy intrinsic alignment has
negligible contamination to  the galaxy-lensing cross correlation measurement, since
the weighted galaxy distribution has vanishing weighting within the
source redshift bin. Additive error is expected to be uncorrelated
with the large scale structure and therefore is not expected to bias
the galaxy-lensing cross correlation measurement.   For these reasons,
the measured $m$ using method 
two is insensitive to neither contamination of galaxy intrinsic
alignment nor additive error in shear estimation. 

In contrast, method one is susceptible to galaxy intrinsic
alignment and additive error. The magnification auto 
  power spectrum contributes little to constraining multiplicative
  error since it suffers from much larger measurement error
  (e.g. Fig. \ref{fig:cl}). Therefore method one basically  interprets the relative difference between the
measured cosmic shear power spectrum and magnification-cosmic shear
cross power spectrum as multiplicative error. However, galaxy intrinsic
alignment and additive error contaminate the cosmic shear power spectrum in
Eq. \ref{eqn:C}, and therefore can cause fake diagnosis of
multiplicative error. 

This potential problem in method one can be rendered into valuable
measurement of galaxy intrinsic alignment/additive error, with the
help of method two.  Basically, method two determines $m$ by the ratio of
galaxy-cosmic shear cross correlation and galaxy-magnification cross
correlation, free of intrinsic alignment/additive error. Method one
measures the combination of $m$ and intrinsic 
alignment/additive error through the ratio of magnification-cosmic shear cross
correlation and cosmic shear auto correlation. With the measured $m$
from method two, we can isolate the combined effect of galaxy
intrinsic alignment and additive error. Therefore, combining method one and
method two, in principle one can measure  galaxy intrinsic
alignment/additive error and multiplicative error simultaneously.

So far we have demonstrated the potential of SN magnification in
calibrating multiplicative error in cosmic shear measurement. There
are a number of caveats in the proposed calibration. One is the
underlying assumption $C^\mu=4C^\gamma$. In reality,
$\mu=1+2\kappa+3\kappa^2+\gamma^2\cdots$. For cosmic shear  we actually
measure the reduced shear
$g=\gamma/(1-\kappa)=\gamma+\gamma\kappa+\cdots$. These high order
terms lead to $C^\mu\neq 4C^\gamma$ and the induced difference in the
two properties is of the order $\sigma^2_\kappa\sim 10^{-3}$. With the
presence of these high order terms, measuring $m$ will rely on
modelling these terms and therefore rely on cosmology. These
complexities can be incorporated by simultaneously fitting $m$ and  
cosmological parameters, which determine these high order
terms. 

Another potential problem is dust extinction by intergalactic gray
dust \citep{Corasaniti06}. Such extinction causes little reddening and
therefore can not be efficiently corrected by conventional reddening recipes. The induced flux fluctuation is spatially correlated and
therefore biases lensing measurement from SN magnification
\citep{Zhang07a}. It is a potential problem for calibrating
multiplicative error with SN magnification. It is also a potential problem for lensing
measurement based on galaxy flux fluctuations
(e.g. \citet{Schmidt12}), and multiplicative error calibration with
galaxy flux fluctuations \citep{Vallinotto11}.  Fortunately, in
principle this problem can be alleviated. Gray dust
extinction induces galaxy number density fluctuation, which differs from that
induced by lensing \citep{YangXJ15}. Therefore we can infer the dust
extinction through galaxy clustering and eliminate it in SN
magnification.  Nevertheless, given large uncertainty in our
understanding of intergalactic gray dust,  it is an important open
question to pay attention.

 Error in  photometry calibration can also potentially cause problem. Its calibration error can be spatially
  correlated. It affects both method one and method two. For method one,
  it directly alters SN flux and biases the  power spectrum
  measurement by lensing magnification. It also
  affects the number of galaxies in bins of observed magnitude.  It
  then indues a correlation between the SN flux fluctuation and galaxy
  number overdensity. Thus it can bias the calibration of
  $m$ using method two.\footnote{We thank the anonymous referee for pointing out this
    issue. } The approach proposed in
  \citet{Vallinotto11} by calibrating multiplicative error with the lensing induced size
bias is free of both the gray dust extinction problem and  the
photometry calibration problem.  So it provides
an independent and highly complementary approach to diagnose and
calibrate multiplicative error.

SNe Ia are highly complementary to other cosmological probes. They not only contribute
as standard candles and valuable measures of weak lensing. They
have already provided robust measurement of peculiar 
velocity at $z\la 0.05$ \citep{Bonvin06,Haugbolle07,Watkins07,Dai11}
and will in the future  even to $z\sim 0.5$ \citep{Zhang08a}. With
millions of SNe Ia, they can be used as 
tracers of large scale structure to measure baryon acoustic
oscillation \citep{Zhan08}.  Our work adds a new application of SNe
Ia, and a new reason to include survey of million SNe Ia by the  LSST
telescope or other weak lensing facilities. 

\section{Acknowledgments}
I thank Jun Zhang and Hu Zhan for helpful discussions. This work was supported by the National
Science Foundation of China 
(Grant No. 11025316, 11320101002, 11433001), National
Basic Research Program of China (973 Program 2015CB857001), the
Strategic Priority Research Program "The Emergence of Cosmological
Structures" of the Chinese Academy of Sciences (Grant
No. XDB09000000), and the key laboratory grant from the Office of Science and Technology, Shanghai Municipal Government (No. 11DZ2260700).

\appendix
\section{Deriving the constraint on multiplicative error using method one}
The Fisher matrix (Eq. \ref{eqn:Fab}) can be decomposed into four blocks,
\be
\label{eqn:FL}
{\bf F}=\bordermatrix{ & & \cr
                &{\bf M}&{\bf E}\cr
                &{\bf G}&{\bf H}\cr}\ .
\ee
${\bf M}\equiv F_{00}$ is in fact a single number, 
\ba
M&=&\sum_{i=1}^n \frac{1}{2}{\rm Tr}\left[{\bf C}^{-1}({\bf
    \ell}_i){\bf C}_{,0}({\bf \ell}_i) {\bf C}^{-1}({\bf
    \ell}_i){\bf C}_{,0}({\bf \ell}_i)\right]\equiv \sum_i M_i\ . \no
\ea
$E=G^T$ is a $1\times n$ matrix, with components
\be
E_i\equiv F_{0i}=\frac{1}{2}{\rm Tr}\left[{\bf C}^{-1}({\bf
    \ell}_i){\bf C}_{,0}({\bf \ell}_i) {\bf C}^{-1}({\bf
    \ell}_i){\bf C}_{,i}({\bf \ell}_i)\right]\ .
\ee
$H$ is a $n\times n$ diagonal matrix, with diagonal elements
\be
H_{ii}=\frac{1}{2}{\rm Tr}\left[{\bf C}^{-1}({\bf
    \ell}_i){\bf C}_{,i}({\bf \ell}_i) {\bf C}^{-1}({\bf
    \ell}_i){\bf C}_{,i}({\bf \ell}_i)\right]\ .
\ee
The inversion of ${\bf F}$ can be done by block operation,
\be
\label{eqn:IF}
\left({\bf F}^{-1}\right)_{00}=\left({\bf M}-{\bf E}{\bf H}^{-1}{\bf
    G}\right)^{-1}\ .
\ee
Since ${\bf H}$ is diagonal, we have 
\be
\label{eqn:reducedA}
{\bf M}-{\bf E}{\bf H}^{-1}{\bf
  G}=\sum_{i=1}^n\left[ M_i-E_iH^{-1}_{ii}G_i\right]\ .
\ee
The error in $m$ is then
\be
\sigma_m=\left[F^{-1}\right]_{00}^{1/2}=\left[\sum_{i=1}^n \left(M_i-E_iH^{-1}_{ii}G_i\right)\right]^{-1/2}\ .
\ee
From Eq. \ref{eqn:C}, we can do the matrix inversion analytically to obtain ${\bf
    C}^{-1}$. We then plug the expression of ${\bf C}$, ${\bf
    C}^{-1}$ and ${\bf C}_{,i}$ into the above equations. Finally we obtain
\ba
\label{eqn:m3}
\sigma_m&=&\left(\sum_i
  \frac{C^{\gamma,2}(\ell_i)}{C^\gamma((1+m)^2N^\mu/4+N^\gamma)+N^\gamma
    N^\mu/4}\right)^{-1/2}\simeq \left(\sum_i
  \frac{C^{\gamma,2}(\ell_i)}{C^\gamma(N^\mu/4+N^\gamma)+N^\gamma
    N^\mu/4}\right)^{-1/2} \ .
\ea
The sum over independent modes ($\sum_i$) can be replaced by the
integral in the continuum limit. Finally we obtain
\ba
\label{eqn:m2}
\sigma_m\simeq \left(\int 
  \frac{2\ell d\ell f_{\rm sky} C^{\gamma,2}(\ell)}{C^\gamma(\ell)(N^\mu/4+N^\gamma)+N^\gamma
    N^\mu/4}\right)^{-1/2} \ .
\ea
This is the most important result for method one, and is used in numerical
evaluations shown in Fig. \ref{fig:m}, \ref{fig:w} and Table
\ref{table:m}.

\bibliographystyle{apj}
\bibliography{mybib}

\end{document}